\documentclass[superscriptaddress,
  aip,
  pop,
  amsmath,amssymb,
  reprint,
]{revtex4-1}
\usepackage{graphicx,color}
\usepackage{amsmath}
\usepackage{natbib}
\usepackage{epsfig}
\begin{document}

\title{{\color{blue} Collective modes of two-dimensional classical Coulomb fluids}}

\author{Sergey A. Khrapak}
\email{Sergey.Khrapak@dlr.de}
\affiliation{Institut f\"ur Materialphysik im Weltraum, Deutsches Zentrum f\"ur Luft- und Raumfahrt (DLR), 82234 We{\ss}ling, Germany}

\author{Nikita P. Kryuchkov}
\affiliation{Bauman Moscow State Technical University, 105005 Moscow, Russia}

\author{Lukia A. Mistryukova}
\affiliation{Bauman Moscow State Technical University, 105005 Moscow, Russia}

\author{Alexey G. Khrapak}
\affiliation{Joint Institute for High Temperatures, Russian Academy of Sciences, 125412 Moscow, Russia}

\author{Stanislav O. Yurchenko}
\email{st.yurchenko@mail.ru}
\affiliation {Bauman Moscow State Technical University, 105005 Moscow, Russia}

\date{\today}

\begin{abstract}
Molecular dynamics simulations have been performed to investigate in detail collective modes spectra of two-dimensional Coulomb fluids in a wide range of coupling. The obtained dispersion relations are compared with theoretical approaches based on quasi-crystalline approximation (QCA), also known as the quasi-localized charge approximation (QLCA) in the plasma-related context. An overall satisfactory agreement between theory and simulations is documented for the longitudinal mode at moderate coupling and in the long-wavelength domain at strong coupling. For the transverse mode, satisfactory agreement in the long-wavelength domain is only reached at very strong coupling, when the cutoff wave-number below which shear waves cannot propagate becomes small. The dependence of the cutoff wave-number for shear waves on the coupling parameter is obtained.         
\end{abstract}

\maketitle

\section{Introduction}

Two-dimensional and quasi-two-dimensional classical interacting particle systems have attracted tremendous interest over the years.~\cite{Kosterlitz1978,KosterlitzRMP2017}
This interest is at least twofold.  First, 2D systems play an important practical role in a broad range of phenomena occurring at fluid and solid surfaces and various interfaces. Examples are atomic monolayers and thin films on substrates, 2D electron fluid on the surface of liquid helium,~\cite{GrimesPRL1979} metallic and magnetic layer compounds, colloidally stabilized emulsions and bubbles,~\cite{PoulichetPNAS2015,YurchenkoLangmuir2016} colloidal particles at flat interfaces,~\cite{Yakovlev2017,Ovcharov2017} complex (dusty) plasma systems in ground-based laboratory conditions.~\cite{ThomasNature1996,FortovUFN2004,MorfillRMP2009} Second, physics in two-dimensions (2D) can be fundamentally different from that in three-dimensions (3D). A celebrated example is related to the nature of the fluid-solid phase transition in 2D.~\cite{KosterlitzRMP2017,RyzhovUFN2017}  

The focus of the present study is on a particular realization of classical 2D systems -- system of point-like charged particles with Coulomb interactions. This system received considerable attention in connection to electron clouds confined in two dimensions,~\cite{BausPR1980,GrimesPRL1979,GannPRB1979} and, more recently, in the context of charged particles in a weakly screened environment (e.g. colloids and complex plasmas).~\cite{FortovUFN2004,FortovPR2005,IvlevBook,ChaudhuriSM2011,YurchenkoPRE2017} Our particular attention will be on the properties of collective modes in the system. Collective modes in 2D Coulomb solids are relatively well understood.~\cite{BonsallPRB1977}  In 2D Coulomb fluids collective modes were actively investigated theoretically in late 70's and 80's.~\cite{OnukiJPSJ1977,BausJSP1978,AgarwalPLA1981,IwamotoPRA1984} Later, the approach known as the quasi-localized charge approximation (QLCA) has been applied.~\cite{GoldenPRA1990,GoldenPRA1992_1}  However, to the best of our knowledge, these theoretical approaches have never been tested thoroughly against experiments and numerical simulations. In fact, we are aware of only one important numerical experiment performed by Totsuji and Kakeya.~\cite{TotsujiKakeyaPLA1979,TotsujiKakeyaPRA1980}    
The limited set of data obtained at that time was not sufficient for a very detailed comparison with theoretical predictions. 

The purpose of this work is to fill this gap and report extensive numerical simulations on the dynamics of 2D Coulomb systems in a wide range of coupling. We have performed molecular dynamics (MD) simulations with $10^4$ particles using the particle-particle particle-mesh (PPPM) Ewald summation method.~\cite{LeBardSoftMatter2012} In this way the dispersion relations of the longitudinal and transverse collective excitations are obtained. These dispersion relations are then compared with the results of theoretical calculations based on the QCA approach. This approach is known to be a good approximation to describe elastic collective modes in strongly coupled fluids with soft long-range interactions (though it fails in the limit of very steep hard-sphere or hard-disk interactions~\cite{KhrapakSciRep2017}). We discuss to which extent it is reliable in the present case of 2D Coulomb fluids. In particular, we demonstrate that QCA should not be applied to short-wavelength excitations and explain why it is so. For the transverse mode, QCA is only meaningful at sufficiently strong coupling  when the critical wave number, below which transverse excitations cannot exist, becomes numerically small. We report the measured dependence of this critical (cutoff) wave number on the coupling parameter.     

Another purpose of the present work is to put the 2D Coulomb system in the context of other classical soft interacting particles in 2D geometry. In a series of papers we have reported detailed investigation of thermodynamics and dynamics of several classical systems of soft-interacting particles in 2D, such as one-component plasma (characterized by logarithmic interaction between the particles),~\cite{KhrapakCPP2016,KhrapakPoP2016_Log} Yukawa systems,~\cite{KryuchkovJCP2017} and dipole-like systems (with $\propto r^{-3}$ interaction).~\cite{KhrapakPRE2018} Present work represents a logical continuation of these studies. The reported results complement those previously obtained and complete the story. 

The paper is organized as follows. In Sec.~\ref{Method} we describe in detail the system under investigation, provide
necessary details about the MD simulations and data analysis, and describe the QCA-based approaches to the collective modes dispersion relations. In Sec.~\ref{Res} main results from MD simulations are presented and compared with theoretical approximations. 
This is followed by our conclusion in Sec.~\ref{Concl}. Some details on the thermodynamic properties and on the derivation of analytical dispersion relations of 2D Coulomb fluids are summarized in Appendices~\ref{A0} and \ref{A1}, respectively. 

\section{Methods}\label{Method}

\subsection{Coulomb systems in two dimensions} 

The Coulomb potential reads
\begin{equation}\label{Coulomb}
\phi(r) = e^2/r,
\end{equation}
where $e$ is the particle charge and $r$ is the interparticle distance. The point-like particles interacting via the potential (\ref{Coulomb}) are immersed into a fixed neutralizing background to stabilize the system and make thermodynamic properties meaningful. The phase behavior of the systems is described by the Coulomb coupling parameter,
\begin{equation}
\Gamma=e^2/aT,
\end{equation}
where $T$ is the temperature (in energy units), $a=(\pi n)^{-1/2}$ is the 2D Wigner-Seitz radius, and  $n$ is the areal (2D) density.  The system is conventionally referred to as strongly coupled when the potential energy [see Eq.~(\ref{uex2}) for the definition] dominates over the kinetic energy, which occurs at $\Gamma\gg 1$. In the opposite limit ($\Gamma \ll 1$) the system is called weakly coupled. 

At very low $\Gamma$ the system properties are similar to those of an ideal gas in 2D. When coupling increases the system forms a strongly coupled fluid phase, which can crystallize into a triangular lattice upon further increase in $\Gamma$.~\cite{GrimesPRL1979} The fluid-solid phase transition occurs at $\Gamma\simeq 120-140$.~\cite{GrimesPRL1979,GannPRB1979} The nature of the fluid-solid phase transition in 2D systems is known to depend considerably on the potential softness.~\cite{KapferPRL2015} For the repulsive power-law interactions ($\propto r^{-\alpha}$) it has been recently demonstrated~\cite{KapferPRL2015} that  for $\alpha\gtrsim 6$ the hard-disk melting scenario holds with the first-order liquid-hexatic transition and a continuous hexatic-solid transition.~\cite{ThorneyworkPRL2017} For $\alpha\lesssim 6$, the liquid-hexatic transition is continuous, with correlations consistent with the Berezinsky-Kosterlitz-Thouless-Halperin-Nelson-Young (BKTHNY) scenario.~\cite{KosterlitzRMP2017} The systems with extremely soft and long-ranged interactions with $\alpha<6$  are not yet accessible for large-scale simulations necessary to establish the exact melting scenario in the thermodynamic limit.~\cite{KapferPRL2015} However, no further change of scenario is expected,~\cite{KapferPRL2015} and the melting of the 2D Coulomb solid should be described by the BKTHNY theory. This important point is beyond the scope of this paper, which is mainly focused on collective modes in the fluid phase.    

Apart from the nature of the fluid-solid phase transition, many other properties of the Coulomb systems in 2D have been studied over decades. Particularly relevant for the present work are the studies of static, dynamical and elastic properties of Coulomb solids;~\cite{BonsallPRB1977,MorfPRL1979} thermodynamic properties of the fluid and solid phases;~\cite{TotsujiPRA1978,TotsujiPRA1979,GannPRB1979,ItohPRA1980,ItohPRB1980,KhrapakPoP2014,KhrapakCPP2016} collective modes of 2D electron fluids;~\cite{OnukiJPSJ1977,BausJSP1978,TotsujiKakeyaPLA1979,TotsujiKakeyaPRA1980,GoldenPRA1990,GoldenPRA1992_1,AgarwalPLA1981,IwamotoPRA1984} and transport properties of the 2D Coulomb fluids.~\cite{HansenPRL1979}         

The considered Coulomb system should not be confused with a similar 2D system in which the interaction potential is defined via the 2D Poisson equation and scales logarithmically with distance.  Both systems are often referred to as the 2D one-component plasma (OCP).  
2D OCP with logarithmic interactions has also received considerable attention,~\cite{JancoviciPRL1981,CaillolJSP1982,LeeuwPhysA1982,AKhrapakAIP2015} including collective modes description,~\cite{LeeuwPhysA1983,KhrapakPoP2016_Log} but this will not be discussed here.   

\subsection{Molecular dynamics simulations}
\label{MDdetails}

We have performed extensive MD simulations of 2D classical Coulomb systems in the $NVT$ ensemble consisting of $N=10^4$ particles. The point-like particles are interacting via the Coulomb potential (\ref{Coulomb}). We have used the PPPM Ewald summation method~\cite{LeBardSoftMatter2012} to account for long-range Coulomb interactions with the cut-off radius of $7.5 n^{-1/2}$ for the short-range part. The system evolves in the Langevin thermostat with a sufficiently low dissipation rate, so that the atomistic dynamics is realized. The numerical time step has been chosen $\Delta t=5.6\times 10^{-4}\sqrt{m a^3\Gamma/e^2}$. All simulations have been performed in a HOOMD-blue package.~\cite{AndersonJCompP2008, GlaserJCompP2015} We have investigated a wide range of coupling parameters corresponding to the fluid phase, $1\leq \Gamma\leq 100$.

\begin{figure}
\includegraphics[width=8cm]{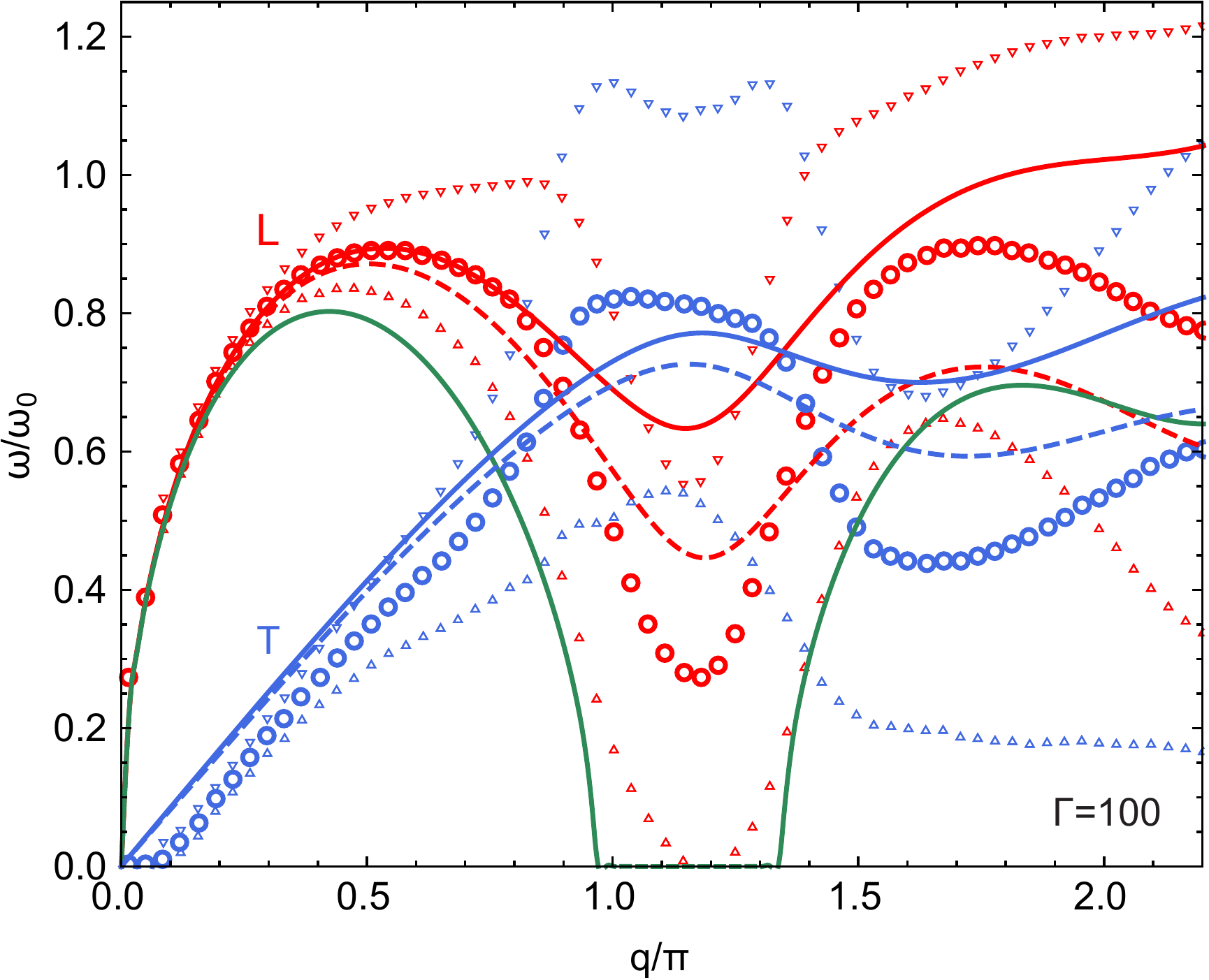}
\caption{Collective modes of the 2D Coulomb fluid with $\Gamma = 100$. Circles correspond to frequencies $\omega_{l,t}$ and triangles to $\omega_{l,t}\pm \gamma_{l,t}$, obtained by applying a fitting function (\ref{wave_ampl_fit}) to the MD data. Red (blue) color corresponds to the longitudinal (transverse) dispersion. The solid and dashed curves of the corresponding color correspond to the QCA approximation with kinetic terms retained, Eq.~(\ref{Disp}), and without kinetic terms, Eqs.~(\ref{DispNoKin}). The green curve corresponds to the generalized high-frequency bulk modulus of Eq.~(\ref{BulkModDisp}). }
\label{Fig1}
\end{figure}

\begin{figure}
\includegraphics[width=8cm]{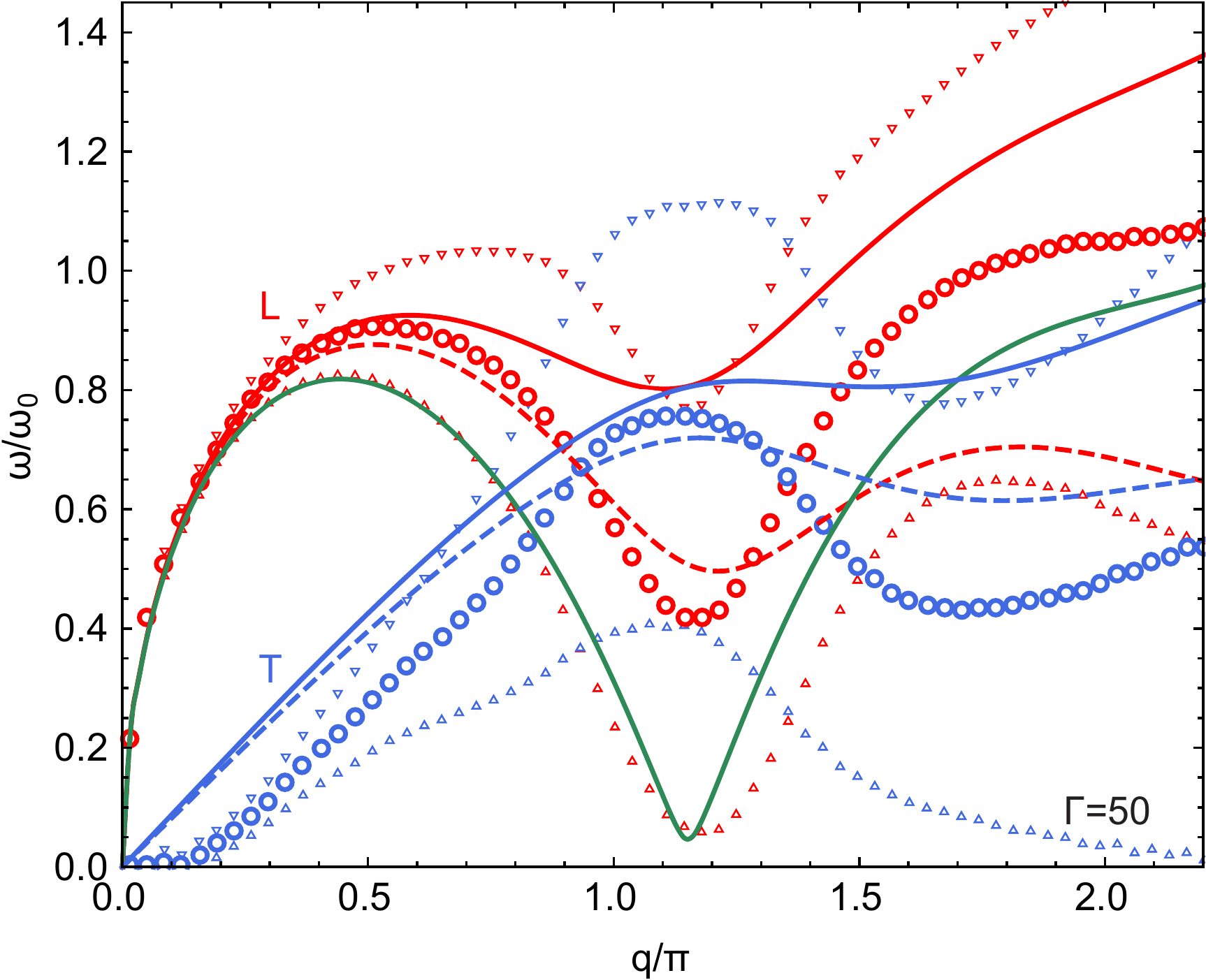}
\caption{Same as in Fig.~\ref{Fig1}, but for the 2D Coulomb fluid with $\Gamma = 50$.}
\label{Fig2}
\end{figure}

The spectra of collective modes in Coulomb fluids have been determined by the standard approach,~\cite{GoldenPRE2010,KhrapakPRE2018, YurchenkoJCP2018} based on measuring the velocity current, 
\begin{equation}\label{vel_current}
 \mathbf{j}(\mathbf{k},t)\propto\sum_s \mathbf{v}_s(t) \exp(i \mathbf{k} \mathbf{r}_s(t)),
\end{equation}
where $\mathbf{v}_s(t)$ and $\mathbf{r}_s(t)$ are the velocity and radius-vector of the $s$-th particles, $ \mathbf{k}$ is the wave vector, and the summation is over all particles in the system. Then we evaluate the longitudinal ($l$) and transverse ($t$) waves amplitudes:
\begin{equation}\label{wave_ampl}
	C_{l,t}(\mathbf{k},\omega) \propto \mathrm{Re}\int dt\, \left<j_{l,t}(\mathbf{k},t)j_{l,t}(\mathbf{-k},0)\right> e^{i\omega t},
\end{equation}
where $j_l(\mathbf{k},\omega)$ and $j_t(\mathbf{k},\omega)$ are projections of velocity current $\mathbf{j}(\mathbf{k},t)$ to the longitudinal and transverse directions, respectively, and $\omega$ is the frequency.  Note, that $C_{l,t}(\mathbf{k},\omega)$ depend only on $k=|\mathbf{k}|$ due to fluid isotropy. To obtain dispersion relations  $\omega_{l,t}(k)$ we fitted $C_{l,t}(k,\omega)$ by the distribution \cite{KhrapakPRE2018, YurchenkoJCP2018}
\begin{equation}\label{wave_ampl_fit}
 f_{l,t}(\omega)\propto\frac{1}{(\omega-\omega_{l,t})^2+\gamma_{l,t}^2}+\frac{1}{(\omega+\omega_{l,t})^2+\gamma_{l,t}^2}
\end{equation}
for each wave-number $k$. Two examples of the obtained dispersion relations in terms of $\omega_{l,t}(k)$ and $\gamma_{l,t}(k)$, corresponding to the strongly coupled regime, are shown in Figures~\ref{Fig1} and \ref{Fig2}.

\subsection{Theory}

A powerful theoretical approach to describe collective motion in classical systems of strongly interacting particles is the quasi-crystalline approximation (QCA).~\cite{Hubbard1969,Takeno1971} This approach can be considered  as either a generalization of the random phase approximation or, alternatively, as a generalization of the phonon theory of solids~\cite{Hubbard1969} (the latter explains why it is often referred to as QCA). In the context of plasma physics an analog of the QCA is known as the quasi-localized charge approximation (QLCA). It was initially proposed as a formalism to describe collective mode dispersion in strongly coupled charged Coulomb liquids in both 3D and 2D~\cite{KalmanPRA1990,GoldenPRA1990,GoldenPRA1992,GoldenPRA1992_1} as well as other related systems (for a review see Ref.~\onlinecite{GoldenPoP2000}). The approach unilaterally links dynamic and structural properties of the systems i.e., it allows to calculate collective modes dispersion relations based on the interaction potential and pair correlations. However, the inverse procedure is poorly understood, and theory of pair correlations reconstruction using the interaction potential and dynamic properties is well developed only for crystals.~\cite{YurchenkoJCP2014,YurchenkoJCP2015,YurchenkoJPCM2016,KryuchkovSM2018}  

The dispersion relation of the longitudinal and transverse modes can be written~\cite{NossalPR1968} as
\begin{equation}\label{Disp}
\begin{split}
\omega_l^2 = \frac{k^2}{mn}\left[{\mathcal K}_{\infty}(k)+{\mathcal G}_{\infty}(k)\right],\\
\omega_t^2 = \frac{k^2}{mn}{\mathcal G}_{\infty}(k),
\end{split}
\end{equation}
where $m$ is the particle mass, and  ${\mathcal K}_{\infty}(k)$ and ${\mathcal G}_{\infty}(k)$ represent the generalized high frequency (instantaneous) bulk and shear moduli (we operate with infinite frequency moduli here; for other purposes finite frequency or ``relaxed'' moduli may be more appropriate~\cite{PuosiJCP2012}).  These can be expressed as
\begin{equation}\label{KG}
\begin{split}
{\mathcal K}_{\infty}=2mnv_{T}^2+\frac{mn}{k^2}\left[{\mathcal L}^2(k)-{\mathcal T}^2(k)\right],\\
{\mathcal G}_{\infty} = mnv_{T}^2+\frac{mn}{k^2}{\mathcal T}^2(k),
\end{split}
\end{equation}
where $v_{T}=\sqrt{T/m}$ is the particle thermal velocity scale (the root-mean-square velocity in 2D is $\sqrt{2T/m}$ ) and the general expressions for the configurational terms ${\mathcal L}(k)$ and ${\mathcal T}(k)$ are  
\begin{equation}\label{LT}
\begin{split}
{\mathcal L}^2(k)=\frac{n}{m}\int\frac{\partial^2 \phi(r)}{\partial z^2} g(r) \left[1-\cos(kz)\right]d{\bf r},\\
{\mathcal T}^2(k)=\frac{n}{m}\int\frac{\partial^2 \phi(r)}{\partial y^2} g(r) \left[1-\cos(kz)\right]d{\bf r},
\end{split}
\end{equation}
where $g(r)$ is the radial distribution function (RDF) and $z=r\cos\theta$ is the direction of the propagation of the longitudinal mode (the particles are confined to the $zy$ plane).

In the present notation, the dispersion relations without the kinetic terms, i.e. 
\begin{equation}\label{DispNoKin}
\begin{split}
\omega_{l}^2={\mathcal L}^2(k), \\
\omega_{t}^2={\mathcal T}^2(k),
\end{split}
\end{equation}
would correspond to the standard QCA (or QLCA) approach.~\cite{GoldenPoP2000} The dispersion relation (\ref{Disp}) with kinetic terms retained follow from the second-frequency-moment sum rules for the current correlation functions.~\cite{AgarwalPLA1981} Despite difference in accounting for the kinetic terms, below we will refer to both approaches as {\it QCA-based}. Explicit expressions for ${\mathcal L}(k)$ and ${\mathcal T}(k)$ are available.~\cite{GoldenPoP2000,AgarwalPLA1981} For completeness they are also summarized in Appendix~\ref{A1}. 

In the QCA-based approaches the dispersion relations are directly and relatively simply expressed in terms of the RDF $g(r)$ and the pair interaction potential $\phi (r)$.~\cite{KhrapakPoP2017_Fingerprints} Only very minor modifications (e.g. related to the presence or absence of the neutralizing background) are required to apply the scheme to various physical systems, characterized by distinct interactions and dimensionality. 
It is this relative simplicity and generality, which have made QCA-based approaches to collective modes particularly popular, although more involved and sometimes more accurate theories (like for instance mode coupling theory~\cite{Gotze1975,Bosse1978_1,Bosse1978_2}) also exist. In addition, in some cases QCA can be further simplified by taking a model RDF, which allows for analytical integration in Eq.~(\ref{LT}) and results in particularly simple fully analytical expressions for the dispersion curves without free parameters. Two-dimensional Coulomb systems represent one of such cases and the corresponding expressions are derived in Appendix~\ref{A1}.

The main idea behind these simplified QCA (sQCA) expressions is as follows. Since the dispersion relations (as well as certain thermodynamic properties) depend on the RDF $g(r)$ only under the integral sign, it is not very unreasonable to assume that a simple model RDF can be constructed, which allows to describe the required integral properties. The model RDF can be  quite different from the actual RDF, it should only capture the essential properties affecting the magnitude of the integrals involved. For very soft long-ranged potentials the contribution from the distant interactions is important, where the fluid RDF exhibits small-amplitude oscillations around $g(r)=1$. At strong coupling the contribution from the short distances is small because the particles cannot approach close to each other due to strong repulsion. Effectively, a correlational hole is formed and $g(r)\simeq 0$ inside this hole. A simplest possible model $g(r)$ satisfying these properties is of the form
\begin{equation}\label{RDF}
g(r)=\theta (r-R),
\end{equation}
where $\theta(x)$ is the Heaviside step function and $R$ is the correlational hole radius (which is of the order of the mean interparticle
separation at strong coupling).  Previously, a similar RDF was employed to analyse the main tendencies in the behaviour of specific heat of liquids and dense gases at low temperatures~\cite{Stishov1980} and to calculate the dispersion relation of Coulomb bilayers and superlattices at strong coupling.~\cite{Golden1993} In the context of QCA approach, an appealing benefit of this simple RDF is that
it allows the analytical integration for certain interaction potentials.  Particularly simple and elegant expressions have been recently derived for Yukawa systems and one-component plasma in 3D~\cite{KhrapakPoP02_2016,KhrapakAIPAdv2017,KhrapakIEEE2018} and one-component plasma with logarithmic interactions in two dimensions.~\cite{KhrapakPoP2016_Log} Somewhat less elegant, but still tractable expressions, have been also derived for the 2D system with dipole-like ($\propto r^{-3}$) interaction.~\cite{KhrapakPRE2018}  In Appendix~\ref{A1} we complement these results by deriving fully analytical dispersion relations for the considered case of 2D Coulomb fluid, see Eqs.~(\ref{L_anal}) and (\ref{T_anal}).

In addition to the dispersion relations arising in the QCA-based approaches, we will also consider a long-wavelength hydrodynamic longitudinal dispersion~\cite{OnukiJPSJ1977,IwamotoPRA1984}
\begin{equation}\label{Hydro}
\omega_l^2 = \omega_0^2 k a+\frac{1}{m}\left(\frac{\partial P}{\partial n}\right)_sk^2,  
\end{equation}
where $\omega_0=\sqrt{2\pi n e^2/ma}$ is the characteristic 2D plasma frequency (note that the reduced wave number  $q=ka$ is also extensively used throughout the paper). The derivative of the pressure $P$ with respect to the density is taken under the condition of constant entropy. The hydrodynamic description applies because 2D Coulomb systems are collisionally dominated.~\cite{OnukiJPSJ1977,TotsujiJPSJ1976} A potential generalization of the long-wavelength hydrodynamic dispersion (\ref{Hydro}) is
\begin{equation}\label{BulkModDisp}
\omega_l^2 = \frac{k^2}{mn} {\mathcal K}_{\infty} (k).
\end{equation}
It will be demonstrated below that the right hand sides of (\ref{Hydro}) and (\ref{BulkModDisp}) are indeed very close to each other (although not identical) across coupling regimes in the long-wavelength limit. It will be also shown that, at moderate coupling, the dispersion relation of the form (\ref{BulkModDisp}) is particularly close to the dispersion relation measured in MD experiment.    
 
\section{Results}\label{Res}

\subsection{Transverse mode}\label{TM}

The standard QCA-based approaches are not very useful to describe the transverse dispersion relations in fluids, because damping effects are not included.~\cite{GoldenPoP2000,KhrapakPRE2018} For example, in Figs.~\ref{Fig1} and \ref{Fig2} clear disagreement between the QCA and MD spectra is observed at short wavelengths. The main reason is that QCA does not take into account effects of anharmonicity, which are responsible, in particular, for damping of collective excitations. Indeed, the particles in fluid are considered within the framework of QCA as ``frozen'' near their equilibrium positions, whose statistics is determined by the actual fluid RDF $g(r)$. Then, the excitation spectra are calculated in the harmonic approximation using perturbation theory for small displacements of particles around equilibrium positions. Account of particles' jumps (important for the physics of fluids) cannot be done within the framework of perturbation theory.~\cite{TrachenkoRPP2016} Anharmonicity is related to the short-range region of the interaction potential, which corresponds to large $q$ in the reciprocal space and results in the observed growing discrepancy between the QCA and MD spectra.

\begin{figure}
\includegraphics[width=8cm]{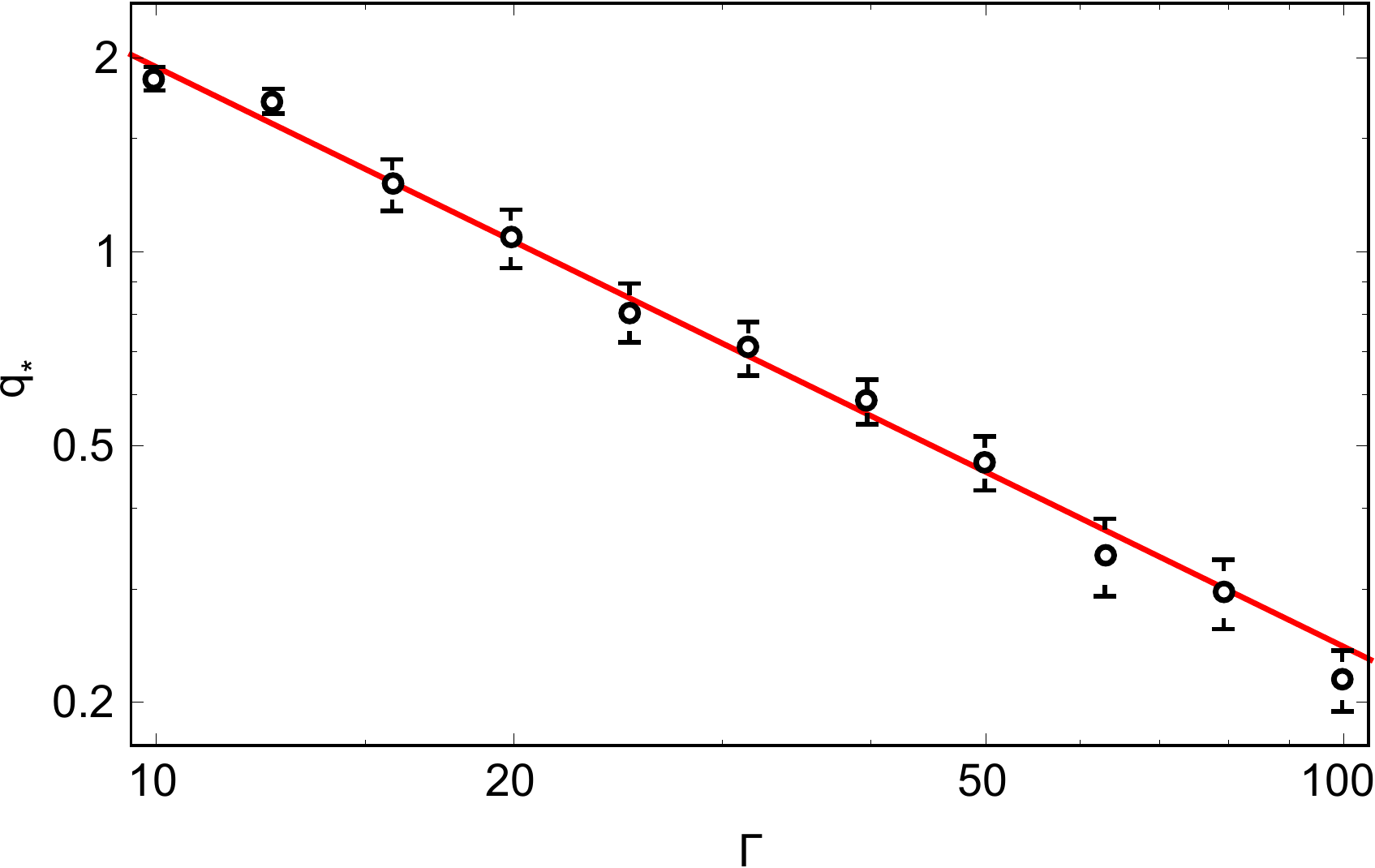}
\caption{Reduced cutoff wave number $q_*$ of the transverse mode in strongly coupled 2D Coulomb fluids versus the Coulomb coupling parameter $\Gamma$. In the strongly coupled regime a decrease of $q_*$ with $\Gamma$ can be reasonably well described by a simple function $q_*\simeq 15.2\Gamma^{-0.9}$, shown by the solid line.}
\label{Fig3}
\end{figure}

Even in the regime of long wavelengths the application of  QCA-based approaches is problematic. In particular, disappearance of the transverse mode at long-wavelengths and the existence of a $q$-gap (minimum wave number $q_*$, also referred to as the ``critical'' or ``cutoff'' wave number, below which shear waves cannot propagate)~\cite{HouPRE2009,GoreePRE2012,YangPRL2017} are not reproduced by theory. From the data presented in Figs.~\ref{Fig1} and \ref{Fig2} we observe that the theory is only relatively close to the numerical data at very strong coupling, $\Gamma\gtrsim 100$, where the $q$-gap is narrow. At lower coupling, the width of the $q$-gap increases with decreasing coupling (weakening correlations) as shown in Fig.~\ref{Fig3}. This behavior is similar to that documented previously for various kinds of simple fluids.~\cite{TrachenkoRPP2016,HouPRE2009,GoreePRE2012,YangPRL2017,SchmidtPRE1997,BrykJCP2017} For 2D Coulomb fluids the dependence of $q_*$ on $\Gamma$ can be fitted by a simple formula  $q_*\simeq 15.2\Gamma^{-0.9}$. 


There exists a simple phenomenological recipe to improve the theoretical description. In the long-wavelength limit,
the generalized hydrodynamic description of the transverse mode yields~\cite{YangPRL2017,OhtaPRL2000} $\omega_t^2\simeq C_t^2k^2-1/(2\tau_r)^2$, where $C_t$ is the transverse sound velocity and  $\tau_r$ is the relaxation time. The condition $q_*\simeq a/(2C_t\tau_r)$ determines the cutoff wave-number. The procedure is then to simply add the term $-1/(2\tau_r)^2$ to the right-hand side of the corresponding theoretical dispersion relation.~\cite{HouPRE2009,KhrapakIEEE2018} In this way, the dispersion relation improves in the long-wavelength low-frequency regime, whilst in the high-frequency regime, where $\omega\tau_r\gg 1$, this correction is negligible. The important real problem of how to estimate the relaxation time from macroscopic or microscopic information available on the system is not yet completely solved (for a recent discussion see e.g. Refs.~\onlinecite{BrykPRL2018,YangPRL2018}). This problem is beyond the scope of the present article. However, the obtained dependence for $q_*(\Gamma)$ can potentially be useful to test various theoretical approximations.   

In the strongly coupled limit, when the width of the $q$-gap diminishes, QCA performance is satisfactory up to the first maximum in the transverse mode dispersion at $q\simeq \pi$ (see Fig.~\ref{Fig1}). In the long-wavelength limit, the transverse sound velocity can be easily related to the system excess energy, see Eq.~(\ref{lowq}). This relation is applicable both in fluid and crystalline phases. In this respect we mention a simple melting criterion of 2D crystals with soft long-ranged interactions proposed recently.~\cite{KhrapakJCP2018} It states that the ratio of the transverse sound velocity of an ideal crystalline lattice to the thermal velocity is a quasi-universal number close to 4.3 at melting.  

In the short-wavelength limit,  the configurational contribution to the transverse dispersion relation is given by the square of the Einstein frequency, $\omega_t^2\simeq \Omega_{\rm E}^2 = (\omega_0^2/2)\int_0^{\infty}g(x)dx/x^2$. This again applies to both liquid and crystalline phases (in the latter case we should sum up instead of integrate). For an ideal crystalline lattice the summation involved represents just the lattice sum for the dipole-dipole ($\propto r^{-3}$) potential.~\cite{KhrapakPRE2018,KhrapakPoP03_2018} For the triangular lattice this yields $\Omega_{\rm E}^2\simeq 0.399256\omega_0^2$.          

\subsection{Longitudinal mode}

It is useful to start with the analysis of the long-wavelength regime. The long-wavelength expansion of the longitudinal dispersion relation reads
\begin{equation}\label{Lwave}
\omega_l^2\simeq \omega_0^2ka+{\mathcal C}k^2v_{T}^2,
\end{equation} 
where ${\mathcal C}$ is the coefficient to be discussed. Theoretically, the first of Eqs.~(\ref{Disp}) yields
${\mathcal C}\simeq 3+\tfrac{5}{8}u_{\rm ex}$, which reduces to the random phase (RPA) approximation $\omega_l^2\simeq  \omega_0^2ka+3 k^2v_{T}^2$ in the absence of correlations. It is known, however, that the mean field approximation is inadequate in the weakly coupled regime of 2D Coulomb systems.~\cite{TotsujiJPSJ1975,BausJSP1978,IwamotoPRA1984} On the other hand, the hydrodynamic description should be appropriate as discusses above, which results in ${\mathcal C}=\gamma\mu$, where $\gamma=c_{\rm P}/c_{\rm V}$ is the adiabatic index and $\mu=(1/T)(\partial P/\partial n)_T$ is the reduced inverse compressibility modulus. The dispersion based on the high-frequency bulk modulus (\ref{BulkModDisp}) implies ${\mathcal C}\simeq 2+\tfrac{3}{4}u_{\rm ex}$ (see Appendix~\ref{A1} for the relation between ${\mathcal C}$ and $u_{\rm ex}$ in different theoretical models). 

It is interesting to compare the predictions of these different models with the actual dispersion relation measured in our numerical experiment. We have, therefore, determined the coefficient ${\mathcal C}$ from MD simulations in a wide range of coupling. We have also calculated ${\mathcal C}$ theoretically using the thermodynamic functions summarized in Appendix~\ref{A0}. The results are plotted in Fig.~\ref{Fig4}. For clarity we show separately the results for moderate (a) and strong (b) coupling regimes. It is observed that at moderate coupling the actual ${\mathcal C}$ is very close to that obtained either from the hydrodynamic dispersion (\ref{Hydro}) or from the high frequency bulk modulus (\ref{BulkModDisp}). The two values are hardly distinguishable from each other. The QCA dispersion relation with kinetic terms (\ref{Disp}) yields significantly higher values of ${\mathcal C}$. Omitting kinetic terms and retaining only the configurational terms [Eq.~(\ref{DispNoKin})] would result in negative values of $C$ for all $\Gamma$, which is clearly irrelevant at moderate coupling, and the corresponding results are not shown in Fig.~\ref{Fig4}(a).  When $\Gamma$ increases, the actual values of ${\mathcal C}$  tend to those predicted by the QCA approach, as observed in  Fig.~\ref{Fig4}(b).
The configurational contribution of Eq.~(\ref{DispNoKin}) is particularly close to MD data. Note however, that kinetic terms are numerically small in this regime, and it is virtually unimportant whether they are retained or not. The hydrodynamic and the high-frequency bulk modulus expression are again practically coinciding, but both slightly underestimate the MD results.   

\begin{figure}
\includegraphics[width=8cm]{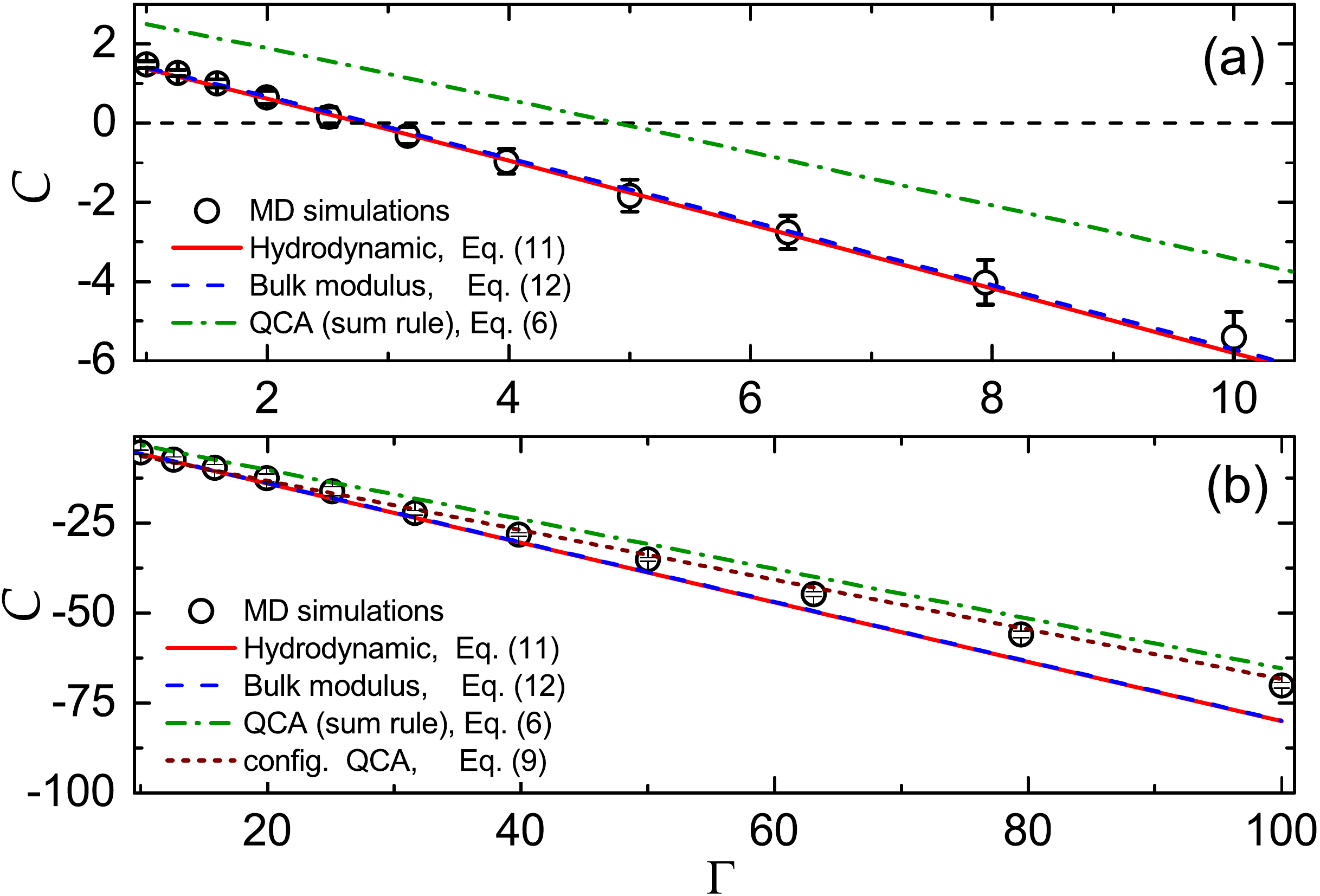}
\caption{The coefficient ${\mathcal C}$ in the long-wavelength expansion of the longitudinal dispersion relation (\ref{Lwave}) versus the coupling parameter $\Gamma$. The top panel (a) corresponds to moderate coupling $1\lesssim \Gamma \lesssim 10$, while  the bottom panel (b) to the strong coupling $10\lesssim \Gamma \lesssim 100$ regime.  Note that ${\mathcal C}$ changes sign from positive to negative near $\Gamma\simeq 3$.}
\label{Fig4}
\end{figure}

In Fig.~\ref{Fig4}(a) we see that ${\mathcal C}$ changes sign from positive to negative at about $\Gamma\simeq 3$.  This is consistent with the observation of Totsuji and Kakeya~\cite{TotsujiKakeyaPLA1979,TotsujiKakeyaPRA1980}  who reported that this change occurs somewhere between $\Gamma=2.29$ and $\Gamma=7.09$. This phenomenon is reminiscent to the onset of negative dispersion in convenient OCP.~\cite{BausPR1980,HansenJPL1981,MithenAIP2012,KorolovCPP2015,KhrapakPoP2016_Onset} The only difference is the character of long-wavelength dispersion, which is $\omega\sim \omega_{\rm p}$ in OCP and $\omega\sim \omega_{\rm p}\sqrt{q}$ in the 2D Coulomb system ($\omega_{\rm p}$ is the plasma frequency, equal to $\omega_0$ in the considered case). It is observed that both the hydrodynamic and the high-frequency bulk modulus approaches are in good quantitative agreement with the simulations results. In the OCP case the conventional hydrodynamic approach is inadequate, because of the high-frequency character of the dispersion.              
On the other hand, the approach based on the analysis of the excess component of the high frequency bulk modulus allows to capture correctly the onset of negative dispersion.~\cite{KhrapakPoP2016_Onset} 

\begin{figure}
\includegraphics[width=8cm]{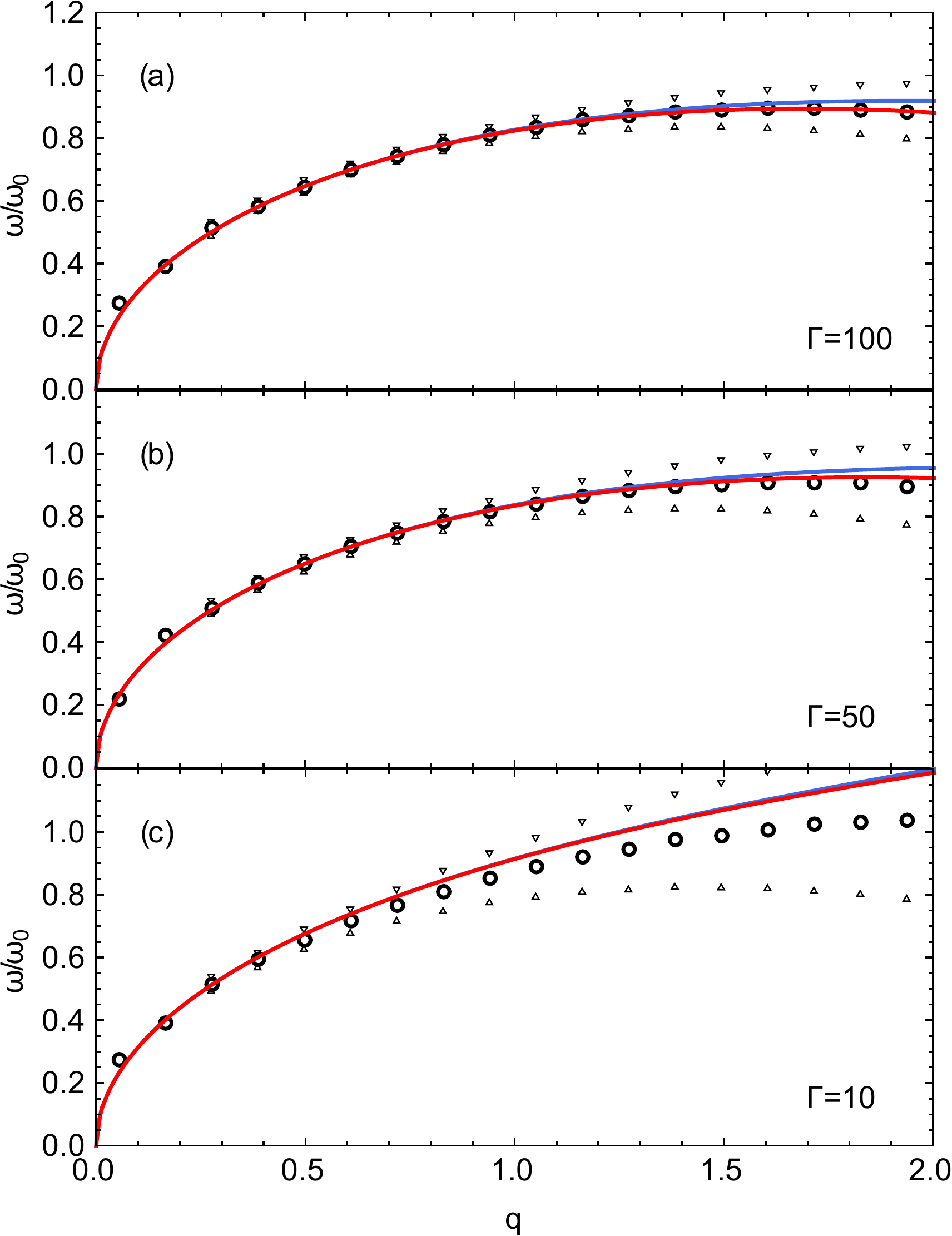}
\caption{Long-wavelength portion of the longitudinal mode dispersion of 2D Coulomb fluid for $\Gamma=100$ (a), $\Gamma=50$ (b) and $\Gamma= 10$ (c). Symbols correspond to the MD simulation data. Red curve are the results of the QCA calculations with the actual RDF. Blue curves are the sQCA calculations using Eq.~(\ref{L_anal}). }
\label{Fig5}
\end{figure}

For shorter wavelengths, the arguments presented in the beginning of Section~\ref{TM} regarding inappropriateness of the QCA approach to describe short-wavelength excitations apply. We observe in Figs.~\ref{Fig1} and \ref{Fig2} that the QCA description agrees with numerical data up to $q\lesssim 0.6\pi$ and is off MD simulation data for higher $q$.
The dispersion relation based on the generalized high-frequency bulk modulus (\ref{BulkModDisp}) is only applicable in the long-wavelength limit, but underestimates considerably the MD frequencies at shorter wavelengths. At strong coupling with $\Gamma=100$, a non-physical region with $\omega_l^2<0$ is observed around $q\simeq \pi$ (see Fig.~{\ref{Fig1}a).  
In the regime where QCA is reliable, the magnitude of kinetic terms is relatively small, which is manifested by closeness of the solid and dashed curves. Moreover, in this regime the fully analytical expression of sQCA, Eqs.~(\ref{L_anal}) and (\ref{T_anal}) are in rather good agreement with the ``full'' QCA approach as well as with the MD data. This is illustrated in Figure~\ref{Fig5}.

\begin{figure*}
\includegraphics[width=17.5cm]{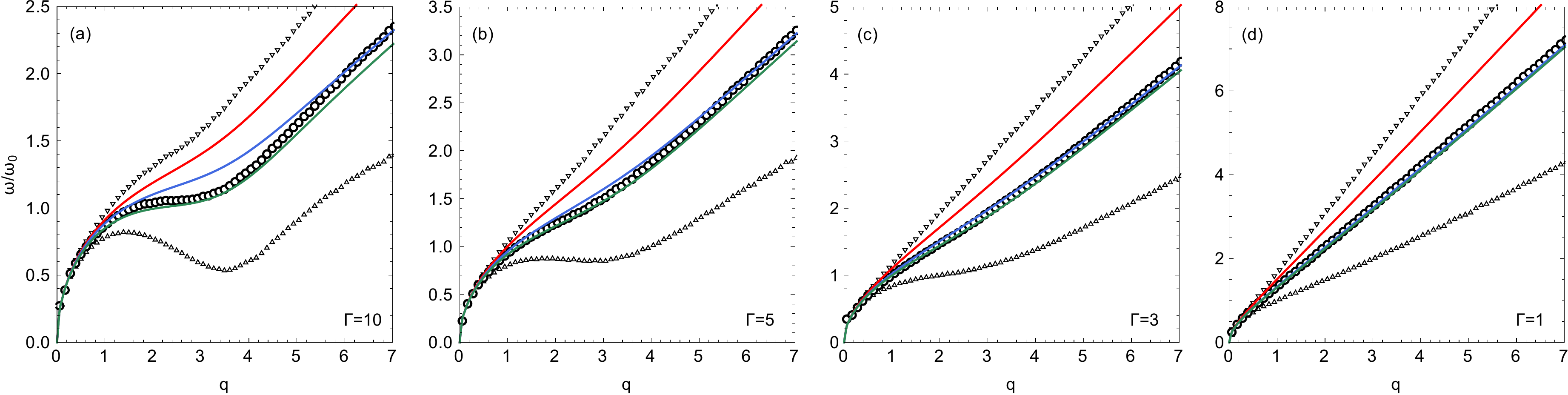}
\caption{Dispersion relation of the longitudinal mode in moderately coupled 2D Coulomb fluids. The results are shown for $\Gamma=10$ (a), $\Gamma=5$ (b), $\Gamma=3$ (c), and $\Gamma=1$ (d). Circles correspond to the main frequency $\omega_l$ and triangles mark the range $\omega_l\pm \gamma_l$, as obtained from MD simulations. The red curves correspond to the QCA approach with kinetic contribution: $\omega_l^2=3k^2v_T^2+{\mathcal L}^2(k)$. The green curves correspond to the generalized high-frequency bulk modulus of Eq.~(\ref{BulkModDisp}). The blue curves shows a phenomenological approach $\omega_l^2=2k^2v_T^2+{\mathcal L}^2(k)$ suggested previously for 2D Yukawa fluids.~\cite{HouPRE2009} }
\label{Fig6}
\end{figure*}

In the moderately coupled regime the dispersion is characterized by the competition between the kinetic and configurational contributions. The main results relevant to this regime are summarized in Fig.~\ref{Fig6}. Good description of the MD results in this regime is provided by the generalized high-frequency bulk modulus, Eq.~(\ref{BulkModDisp}).  Alternatively, one can combine the QCA configurational term ${\mathcal L}(k)^2$ with the 2D ideal gas kinetic term $2k^2v_T^2$. Such a phenomenological approach has been previously proposed in the context of collective modes in 2D dusty plasmas with Debye-H\"uckel (Yukawa) interaction.~\cite{HouPRE2009}       

\section{Conclusion}\label{Concl}

Using the MD simulations we have invetsigated the collective modes dispersion of 2D Coulomb fluids in a wide regime of coupling, from $\Gamma= 1$ to $\Gamma=100$. The obtained new results have been critically compared with theoretical approaches based on the quasi-crystalline approximation.   

Our main results can be shortly summarized as follows. QCA approach is a simple and useful theoretical tool to describe long-wavelength portions of the longitudinal and transverse dispersion relations of 2D Coulomb fluids at strong coupling. For transverse waves this applies only to the very strong coupling regime, where the $q$-gap becomes sufficiently narrow. In this strongly coupled regime good agreement with the MD results is observed at $q\lesssim 0.6\pi$ for the longitudinal mode and $q\lesssim \pi$ for the transverse mode. In this range of wave-numbers a simplified QCA with the step-wise RDF $g(r)$ results in fully analytical parameter-free formulas, which are in good agreement with MD results as well. In this domain the dispersion relations are dominated by configurational terms, kinetic terms can be omitted. The dependence of the cutoff wave-number $q_*$, below which shear waves cannot propagate, on the coupling parameter $\Gamma$ has been obtained.  

In the regime of moderate coupling, the longitudinal mode dispersion relation is characterized by the competition between the kinetic and configurational contributions. The dispersion relation based on the generalized high-frequency bulk modulus is consistent with the simulation results. The kinetic term that dominates the dispersion of weakly coupled 2D Coulomb fluid at short wavelengths is $2k^2v_T^2$, different from the Bohm-Gross term $3k^2v_T^2$, occurring in OCP systems (in 3D and 2D).            

The obtained results complement and improve previously reported results on the dynamical properties and collective modes in 2D Coulomb systems and, more generally, in 2D classical systems with soft long-ranged interactions.    

\acknowledgments

We would like to thank Ingo Laut for a critical reading of the manuscript. MD simulations performed at BMSTU were supported by the Russian Science Foundation, Grant No. 17-19-01691.

\appendix

\section{Thermodynamic functions of 2D Coulomb fluids}\label{A0}

All required thermodynamic functions can be expressed in terms of the dependence of the reduced excess (configurational) energy $u_{\rm ex}$ on $\Gamma$. The integral equation for the reduced excess energy in case of the 2D Coulomb system with the fixed neutralizing background is
\begin{equation}\label{uex2}
u_{\rm ex}=\frac{n}{2T}\int\phi(r)h(r)d{\bf r}=\Gamma\int_0^{\infty}h(x)dx,
\end{equation}
where $h(r)=g(r)-1$ and $x=r/a$.

When $u_{\rm ex}(\Gamma)$ is known, other thermodynamic quantities are obtained as follows. The reduced system energy per particle $u=U/NT$ is
\begin{equation}
u=1+u_{\rm ex}.
\end{equation}
The compressibility $Z=PV/NT$ (reduced pressure) is
\begin{equation}
Z=1+p_{\rm ex}=1+\frac{1}{2}u_{\rm ex}.
\end{equation}
The inverse reduced isothermal compressibility modulus $\mu=(1/T)(\partial P/\partial n)_T$ is
\begin{equation}
\mu = 1+\frac{1}{2}u_{\rm ex}+\frac{\Gamma}{4}\frac{\partial u_{\rm ex}}{\partial \Gamma}.
\end{equation}
The reduced isochoric heat capacity $c_{\rm V}=(1/N)(\partial U/\partial T)_V$ is
\begin{equation}
c_{\rm V}=1+u_{\rm ex}-\Gamma \frac{\partial u_{\rm ex}}{\partial \Gamma}.
\end{equation}
The adiabatic index $\gamma = c_{\rm P}/c_{\rm V}$ of the 2D Coulomb fluid is 
\begin{equation}
\gamma = 1+\frac{(c_{\rm V}+1)^2}{4\mu c_{\rm V}}.
\end{equation}
The product $\gamma\mu$ is 
\begin{equation}
\gamma\mu=\mu +\frac{(c_{\rm V}+1)^2}{4c_{\rm V}}.
\end{equation}
 In the weakly coupled regime ($\mu\simeq 1$, $c_{\rm V}\simeq 1$) we get $\gamma \mu\simeq 2$ as expected for 2D geometry. In the strongly coupled regime ($\mu\gg 1$, $\gamma\simeq 1$) we get $\gamma\mu \simeq \mu$. 

As for the dependence $u_{\rm ex}(\Gamma)$, we have used a simple two-term expression~\cite{KhrapakCPP2016}
\begin{equation}\label{uex1}
u_{\rm ex} =M\Gamma +0.231\ln(1+2.798\Gamma),
\end{equation}
where $M\simeq -1.1061$ is the Madelung constant (triangulat lattice sum) of a 2D Coulomb solid. This functional form has been proven to be very useful for various fluids characterized by soft long-ranged interactions in 2D.~\cite{KhrapakPRE2018,KhrapakPoP08_2015,SemenovPoP2015,KryuchkovJCP2017} 

\section{Simplified expressions for ${\mathcal L}$ and ${\mathcal T}$}\label{A1}

The explicit expressions for  ${\mathcal L}(k)$ and ${\mathcal T}(k)$ for the 2D Coulomb system are~\cite{GoldenPoP2000,AgarwalPLA1981} 
\begin{equation}\label{LandT}
\begin{split}
{\mathcal L}^2(k)=\omega_0^2q + \omega_0^2\int_0^{\infty}\frac{h(x)}{2x^2}\left[1-J_0(qx)+3J_2(qx)\right]dx,\\
{\mathcal T}^2(k)=\omega_0^2\int_0^{\infty}\frac{h(x)}{2x^2}\left[1-J_0(qx)-3J_2(qx)\right]dx,
\end{split}
\end{equation}
where $J_{\alpha}(x)$ are Bessel functions of the first kind, $q=ka$, and $x=r/a$. In the long-wavelength limit ($q\rightarrow 0$) series expansion up to ${\mathcal O}(q^2)$ terms yields 
\begin{equation}\label{lowq}
\begin{split}
{\mathcal L}^2(q)=\omega_0^2q + \frac{5}{8}k^2v_T^2u_{\rm ex},\\
{\mathcal T}^2(q)=-\frac{1}{8}k^2v_T^2u_{\rm ex}.
\end{split}
\end{equation}

For a simplified RDF accounting for a correlational hole at short interparticle separations and absence of correlations at long separations, $g(x)=\theta(x-R)$ ($R$ is now expressed in units of $a$), the integration in Eq.~(\ref{LandT}) can be done analytically, resulting in 
\begin{equation}\label{L_anal}
\begin{split}
\frac{{\mathcal L}^2(q)}{\omega_0^2}=q+\frac{1}{2R}\\ +\frac{J_1(qR)}{2qR^2}\left[2+2q^2R^2-\pi q^3R^3H_0(qR)\right]\\
-\frac{J_0(qR)}{2R}\left[2+2q^2R^2-\pi q^2R^2H_1(qR)\right]
\end{split}
\end{equation}
and
\begin{equation}\label{T_anal}
\frac{{\mathcal T}^2(q)}{\omega_0^2}=\frac{1}{2R}-\frac{J_1(qR)}{qR^2}.
\end{equation} 
Here $H_0(x)$ and $H_1(x)$ denote the Struve functions of order 0 and 1, respectively.  

The reduced correlational hole radius $R$ is not a free parameter of the model. It should be determined from the condition that the energy or pressure integral equations yield adequate results when the model RDF is substituted.~\cite{KhrapakPoP02_2016} For inverse power potentials the energy and pressure routes are equivalent. 
Substituting the model RDF in Eq.~(\ref{uex2}) we immediately get 
\begin{equation}
R=-u_{\rm ex}/\Gamma.
\end{equation}
$R$ is then determined with the help of (\ref{uex1}). In the strongly coupled regime the static (Madelung) contribution to the excess energy is dominant and we obtain $R\simeq 1.1061$. Closer value of $R=\sqrt{6/5}\simeq 1.095$ was previously obtained for the strongly coupled 3D Coulomb system.~\cite{KhrapakPoP02_2016} In the 2D OCP with logarithmic interaction the pressure equation yields~\cite{KhrapakPoP2016_Log} $R=1.0$.          
  
Alternatively, to obtain $R$ at strong coupling we can analyze the long-wavelength expansion of Eqs.~(\ref{L_anal}) and (\ref{T_anal}) which read
\begin{equation}\label{QCA_SC}
\begin{split}
{\mathcal L}^2(q)\simeq {\omega_0^2}\left( q-\frac{5Rq^2}{16}\right), \\
{\mathcal T}^2(q)\simeq {\omega_0^2}\frac{Rq^2}{16}. 
\end{split}
\end{equation}
We then compare these expansions with the corresponding expansions of Bonsall and Maradudin~\cite{BonsallPRB1977} for the low-$q$ limit of collective modes in the ideal hexagonal lattice (this is appropriate because the QCA approach reduces to the conventional phonon theory of solids when applied to an ideal crystalline lattice),
\begin{equation}\label{Bonsall}
\begin{split}
\omega_l^2\simeq {\omega_0^2}\left(q- 0.345657q^2 \right), \\
\omega_t^2\simeq {\omega_0^2}(0.06913q^2). 
\end{split}
\end{equation}
Comparing (\ref{QCA_SC}) and (\ref{Bonsall}) we recover $R\simeq 1.1061$, as should be expected. 

\bibliographystyle{aipnum4-1}
\bibliography{OCP_2D}

\end{document}